# Complex Skin Modes in Non-Hermitian Coupled Laser Arrays

Yuzhou G. N. Liu[1], Yunxuan Wei[1], Omid Hemmatyar[1], Georgios G. Pyrialakos[2], Pawel S. Jung[2,3], Demetrios N. Christodoulides[2], Mercedeh Khajavikhan[1,4,*]

[1]Ming Hsieh Department of Electrical and Computer Engineering, University of Southern California, Los Angeles, California 90089, USA

[2]CREOL, The College of Optics & Photonics, University of Central Florida, Orlando, Florida 32816–2700, USA

[3]Faculty of Physics, Warsaw University of Technology, Koszykowa 75, 00-662 Warsaw, Poland

[4]Department of Physics & Astronomy, Dornsife College of Letters, Arts, & Sciences, University of Southern California, Los Angeles, California 90089, USA

*Corresponding author: khajavik@usc.edu

**Abstract**
From biological ecosystems to spin glasses, connectivity plays a crucial role in determining the function, dynamics, and resiliency of a network. In the realm of non-Hermitian physics, the possibility of complex and asymmetric exchange interactions ($|\kappa_{ij}| \neq |\kappa_{ji}|$) between a network of oscillators has been theoretically shown to lead to novel behaviors like delocalization, skin effect, and bulk-boundary correspondence. An archetypical lattice exhibiting the aforementioned properties is that proposed by Hatano and Nelson in a series of papers in late 1990s. While the ramifications of these theoretical works in optics have been recently pursued in synthetic dimensions, the Hatano-Nelson model has yet to be realized in real space. What makes the implementation of these lattices challenging is the difficulty in establishing the required asymmetric exchange interactions in optical platforms. In this work, by using active optical oscillators featuring non-Hermiticity and nonlinearity, we introduce an anisotropic exchange between the resonant elements in a lattice, an aspect that enables us to observe the non-Hermitian skin effect, phase locking, and near-field beam steering in a Hatano-Nelson laser array. Our work opens up new regimes of phase-locking in lasers while shedding light on the fundamental physics of non-Hermitian systems.

**Introduction**
It is well known that two identical elastic bodies must exchange energy at the same rate as long as the conservative collision process is reciprocal. This situation naturally arises in Hermitian Hamiltonian systems where two parties can exchange energy in a fully symmetric fashion. Interestingly, more than two decades ago, Hatano and Nelson predicted the emergence of delocalization in a special class of exponentiated non-Hermitian random quantum Hamiltonians in order to address some outstanding problems in classical statistical mechanics[1–3]. Such Hamiltonians are quite ubiquitous in nature, describing a wide span of phenomena ranging from various non-equilibrium processes[4–25] to asymmetric XXZ spin chains and spatial inhomogeneities in biological networks[26,27], to name a few. Recent studies also suggest that such lattices provide a route for realizing a new category of topologically non-trivial states in non-Hermitian systems[4,6,28].

The Hatano-Nelson model features a lattice with asymmetric coupling terms induced by an external field that operates like an imaginary vector potential. A schematic of a 1D Hatano-Nelson array along with a conceptual realization in the optical domain are shown in Fig. 1a. The Hamiltonian of this lattice is described by:

$$\hat{H} = -t/2 \sum_n (\exp(-g) \hat{a}^\dagger_{n+1} \hat{a}_n + \exp(g) \hat{a}^\dagger_{n-1} \hat{a}_n) \tag{1}$$

where $\hat{a}^\dagger_n$ and $\hat{a}_n$ are the bosonic creation and annihilation operators at sites $n$; $t$ is the hopping strength, and $g \in \mathbb{R}$ represents a "non-Hermitian external field" (Fig. 1b). In 2D, the Hatano-Nelson model reduces to a quantum Hall system for spatially varying imaginary values of $g$. When subject to periodic boundary conditions, the Hatano-Nelson lattice supports a set of delocalized states with pair-wise complex eigenvalues (see Fig. 1c). On the other hand, under open boundary conditions the eigenvalues are entirely real, whereas the eigenmodes exhibit non-Hermitian skin effects and the energy distribution in the array tilts towards one of the two ends (see Fig. 1d). The directionality and strength of the power imbalance across the array is dictated by the sign and value of the field parameter $g$. Recently, this type of lattice has been demonstrated in acoustic systems[10], mechanical metamaterials[29,30] and electrical circuits[31,32], and also proposed in elastic media[31] and cold atoms[20]. In optics, such lattices have been only implemented in synthetic dimensions, leading to the observation of light funneling with interface localization[12], non-Hermitian bands with arbitrary winding numbers[13], and complex-energy braiding[14]. However, the realization of such non-reciprocal coupling processes in real space have so far remained difficult if not elusive[34–36].

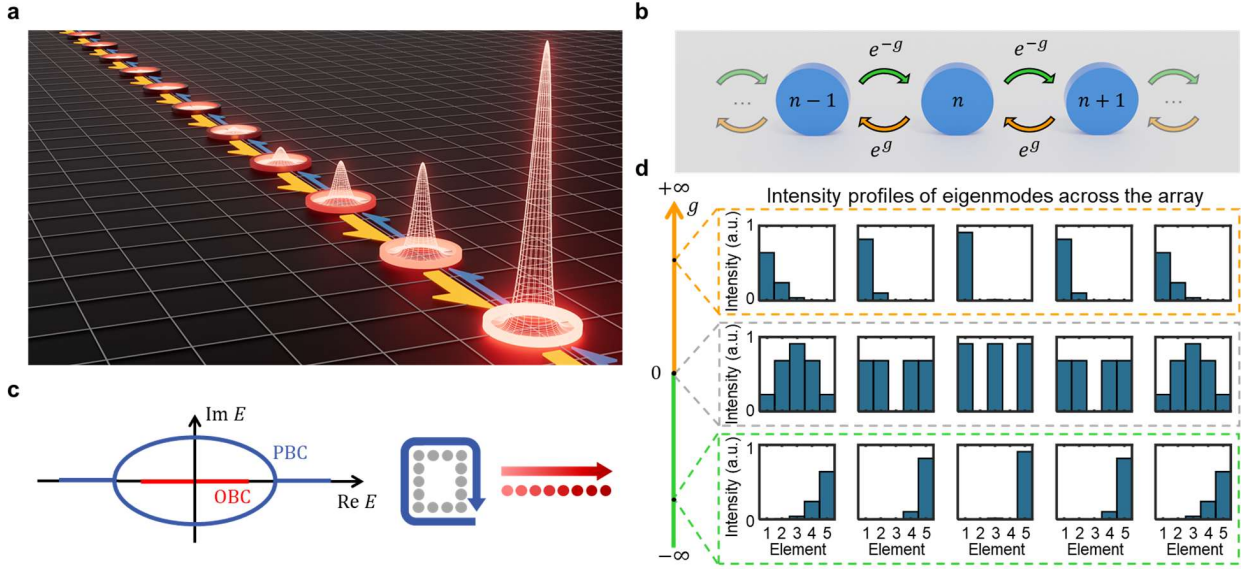

**Fig. 1 | Hatano-Nelson model. a,** A schematic of a 1D Hatano-Nelson lattice. The non-Hermitian skin effect localizes the intensity at the edge of the lattice. **b,** A generalized Hatano-Nelson lattice with asymmetric coupling marked by orange and green arrows. **c,** Eigen-energies of a Hatano-Nelson lattice when periodic boundary condition (PBC) or open boundary condition (OBC) is applied. **d,** Non-Hermitian skin effect appears when the coupling is asymmetric (orange and green sections) thus the energy distribution in the modes tilts towards left and right ends, respectively.

In this work, we utilize non-symmetric exchange interactions arising in judiciously coupled active resonators in order to tailor the response of a corresponding Hatano-Nelson phased locked laser array. The building block of our system is shown in Fig. 2a, where two III-V compound semiconductor active microring resonators are coupled through a pair of asymmetrically terminated link waveguides. These connecting sections provide an asymmetric delayed mutual coupling by reflecting the field at one end (sharply terminated), while gradually dissipating the energy at the tapered side. This coupling scheme affects the spontaneous emission by modifying the density of states, thus promoting an energy circulation in the rings only in one direction[37] (here counter-clockwise) as shown in Fig. 2a. The mathematical description of this behavior is provided in the Supplementary Section 1. The coupling from resonator ① → ② is given by $\kappa_R = i\gamma_u e^{i\beta L}$, whereas that from ② → ① is expressed by $\kappa_L = i\gamma_l e^{i\beta L}$, where $\beta$ is the propagation constant of the TE$_0$ waveguide mode and $L$ is the length of the links. The right-left coupling strengths between the two rings are determined by the gain/loss coefficients, $\gamma_u$ and $\gamma_l$, that are in turn controlled by the pumping profiles of the upper and lower links, respectively. It should be noted that the resulting asymmetric energy exchange between the resonators is enabled in part by the unidirectional circulation of power in the rings[16–18], something that is impossible to achieve in passive structures, even if the links endure varying levels of loss[37].

## Results

While the ratio between $\gamma_u$ and $\gamma_l$ determines the degree of asymmetry between the coupling coefficients ($g = 1/2 \ln(\gamma_l/\gamma_u)$), the value of $\beta L$ plays an important role in governing the lasing properties of the array. When $\beta L = m\pi$, where $m$ is an integer, both coupling coefficients are imaginary, resulting in a splitting in the imaginary parts of the corresponding eigenfrequencies. The two modes of the system will then occur at the same wavelength albeit with different quality factors. Clearly, in this case, the mode with the higher quality factor is poised to lase. This situation is favorable for phase locking as it results in single mode lasing operation. Furthermore, depending on whether the integer $m$ is even or odd, this mode can be in-phase or $\pi$-out-of-phase. It should be noted that the situation described here, where the asymmetric coupling is complex, is unique to our experimental system, and thus represents a generalization of the original Hatano-Nelson model in which under open boundary conditions all eigenvalues are real. On the other hand, when $\beta L = (m + 1/2)\pi$, the coupling coefficients are real, and therefore two lasing modes with identical quality factors will emerge. For any values in between, the system is expected to support two modes at two different frequencies and with varying quality factors. In practice, however, as long as $\beta L$ remains in the vicinity of $m\pi$, the two eigenvalues are complex with a substantial difference between their imaginary components, and therefore single mode lasing is expected to prevail.

The response of the asymmetrically coupled microring lasers is experimentally characterized by examining their emission properties. The aforementioned coupled resonant systems are fabricated on an InP semiconductor wafer, covered by 6 quantum wells of InGaAsP with an overall thickness of 200 nm. The fabrication procedure is outlined in Supplementary Section 2. The ring resonators have a radius of 5 μm. All waveguiding sections feature a high-contrast core ($n_{core} = 3.4$) with a width of 500 nm and a height of 200 nm that is embedded in a silicon dioxide film ($n_{SiO_2} = 1.45$) and is exposed to air on top. These structures are designed to support the TE$_0$ mode with an effective index of $n_{eff} = 2.24$. To promote lasing in a single mode, $\beta L$ of the upper and lower links are designed to be close to $m\pi$ at the operating wavelength (see Supplementary Section 3 for more detail). The fabricated samples are then tested in a $\mu$-photoluminescence setup at room temperature with a pulsed pump laser (wavelength: 1064 nm, pulse width: 15 ns, repetition rate: 290 kHz). To establish the asymmetric coupling, the pump profile is shaped using a combination of knife edges before being imaged on the sample plane, where it is partially blocked from the upper or lower links depending on the sign of asymmetry. In order to confirm the light direction of circulation, each ring is accompanied by a bus waveguide that is terminated at two grating couplers. In the experiments throughout this study, all resonators are uniformly pumped. For more information about the measurement station and methodology see Supplementary Section 4.

Figures 2b and c show the measured emission profiles and spectra of the asymmetrically coupled two-resonator system with $\beta L \cong m\pi$. This phase condition is verified by 16 samples with varying length $L$ (see Supplementary Section 3 for more detail). For the purpose of visual comparison, the layout of the structure is inserted in the background, and the pumped area is specified by a bright rectangle. In this configuration, pumping the upper link leads to $g < 0$ ($g = -1.66$ in Fig. 2b), while $g > 0$ when the lower link is illuminated ($g = 1.66$ in Fig. 2c). Examining the emission intensity from the gratings (which is expected to be linearly proportional to that in the two counter propagating directions) confirms that indeed the energy circulates in a counter-clockwise direction in the rings. In addition, a significant intensity imbalance is observed between the two resonators (attributed to the coupling asymmetry), leading to an energy shift in the array towards the left or right ring as a result of pumping the lower or upper link, respectively. Furthermore, in both cases the emission spectra are single-moded, indicating that the proper coupling phase conditions are experimentally established. Notice that, in Fig. 2b, the bus waveguides are partially pumped in order to reduce the loss in the path towards the gratings, which causes a small power residual appears at the clockwise output. On the other hand, in Fig. 2c, the lower links are not fully pumped, which reduce the spontaneous emission in the bus-waveguides and maintain a relatively high visibility. The light-light curve displayed in Fig. 2d and the spectrum evolution presented in Fig. 2e further confirm that the array indeed operates as a phased-locked laser system.

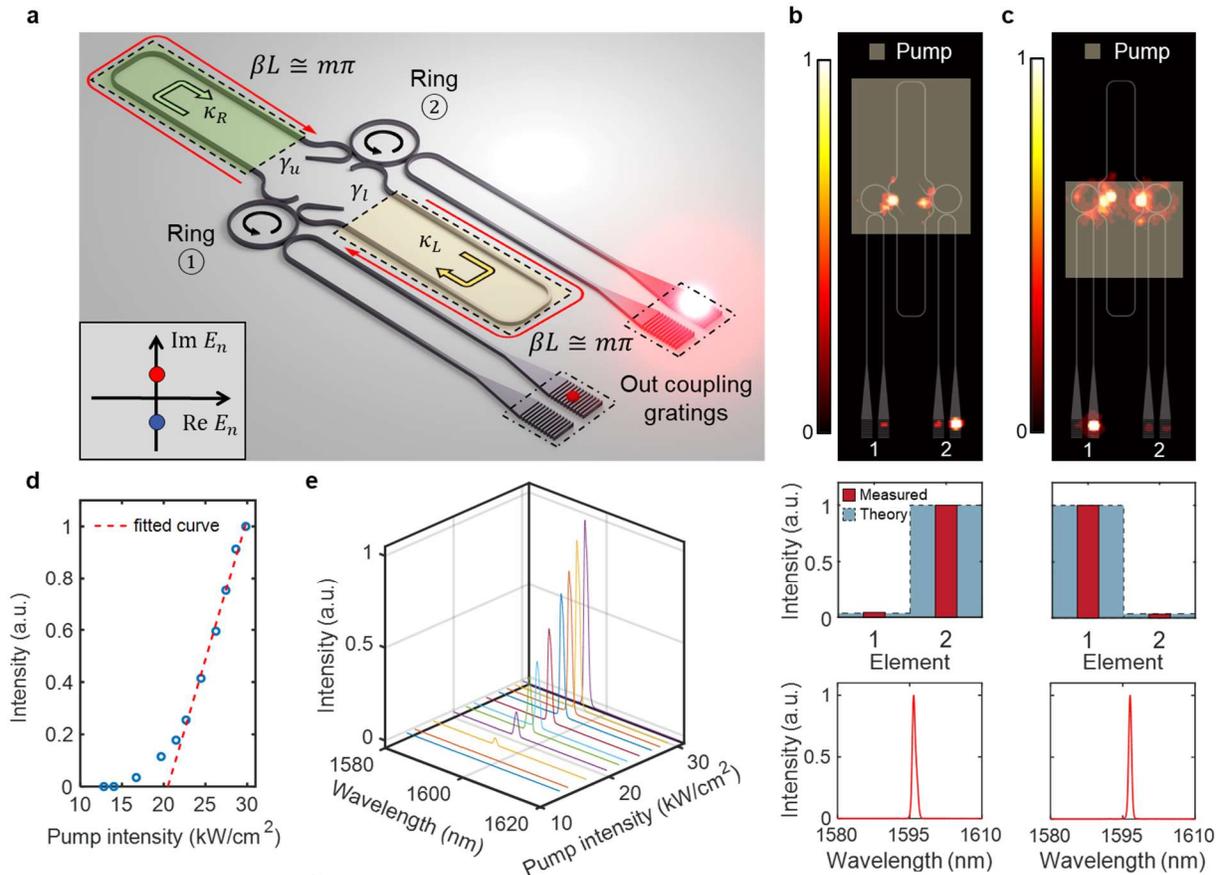

**Fig. 2 | Hatano-Nelson microring laser – 2-element system. a,** Schematic of the unidirectional microring laser with adjustable asymmetric coupling. Lasing in the counter-clockwise direction in each microring is promoted by properly terminating the link structure as to minimize the feedback to the clockwise mode. The magnitudes of the coupling coefficients depend on the pumping profiles of the upper and lower links, $\gamma_u$ and $\gamma_l$, respectively. The length of the links is set to satisfy $\beta L \cong m\pi$. Inset, the eigenvalues of the system become imaginary when the phase of the coupling regions is set at $\beta L \cong m\pi$. In a laser system, the mode with the highest imaginary part eigenenergy (red circle in the figure) will reach lasing threshold first and the laser array operates in a single-mode manner. **b&c,** Emission profiles and spectra of the asymmetrically coupled two-resonator system. Under two different pumping profiles, the non-Hermitian skin effect appears towards two different directions (right and left). The number of data points in both spectra is 432. **d&e,** Light-light curve and spectrum evolution of the laser system when it is operating as in **b**. The resolution of the spectrometer is $0.64\ nm$.

Next, we examine the emission properties of a 5-element Hatano-Nelson laser array, having the above asymmetrically coupled two-resonator system as a building block. Figure 3a depicts the microscope and scanning electron micrograph (SEM) images of the fabricated array. Similar to a two-level system, the coupling ratios can be adjusted by spatially varying the pump profile overlap with the links. Figures 3b, e, and h show the emission intensity of the array when the non-Hermitian field parameter changes from $g = -1.78$ to $g = 0$ to $g = 1.78$. Consequently, the corresponding mode profiles are reported in Figs. 3c, f, and i, where the peak of the lasing mode shifts from one end of the array to the other. As expected, at $g = 0$ the array supports a fully symmetric mode. By gradually shifting the pump profile from covering the upper to the lower link, one can steer the beam in the near field from one end of the array to the other. This situation is recorded in Supplementary Section 5 and Supplementary Videos 1 and 2. In all these cases, the system operates in a single longitudinal mode (Figs. 3d, g, and j) dictated by the mode discrimination afforded by the proper link length choice ($\beta L \approx m\pi$). Similar results are obsreved in larger lattices of 11 elements (see Supplementary Section 6).

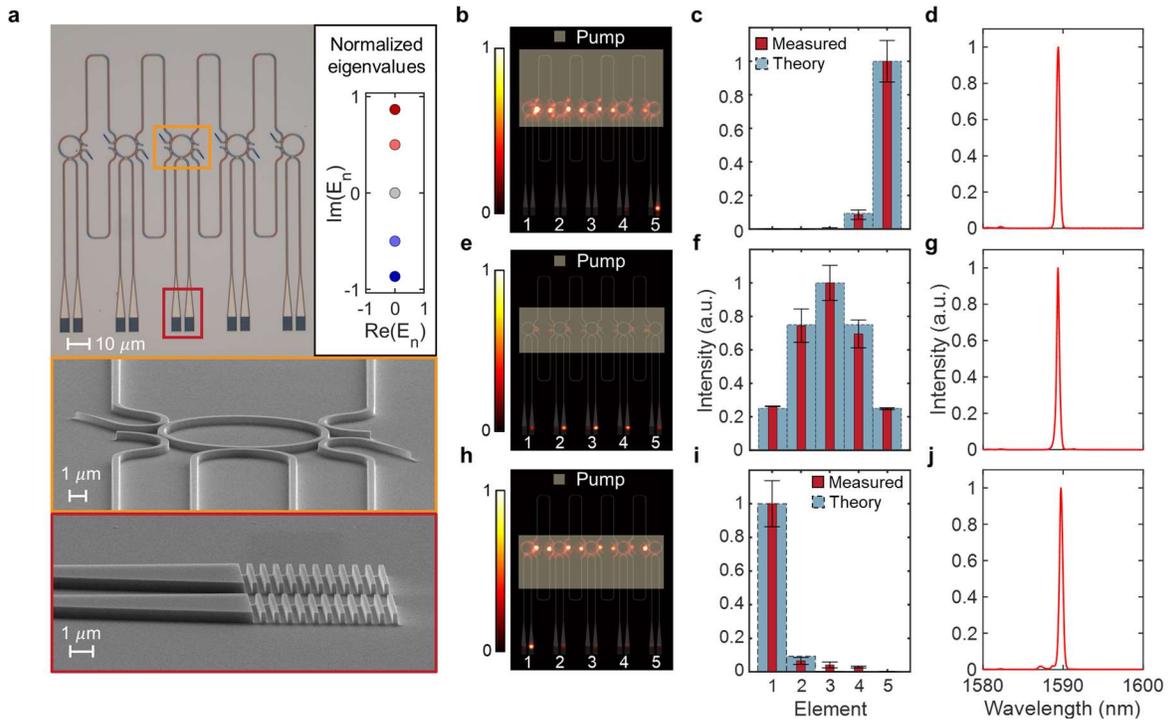

**Fig. 3 | A 5-element Hatano-Nelson phase locked microring laser array. a,** Microscope and scanning electron microscope (SEM) images of the 5-element systems. Inset on the top right, Normalized eigenvalues when $\beta L \cong m\pi$. The mode with highest imaginary eigenvalues (dark red circle) will be the only lasing mode in the system. **b-j,** Experimental conditions and results for the 5-element Hatano-Nelson lattice. Under three different pumping profiles, upper (**b**), middle (**e**), and lower (**h**), the 5-element system can be controlled to demonstrate the non-Hermitian skin effect towards two different directions (**b&h**) or return to its trivial form where the right and left couplings are equal (**e**). The intensity at various sites are reported (the error bars show the variation over 15 measurements) (**c, f&i**). The link length is designed to promote single-mode lasing behavior and the measured spectra confirm that this condition is satisfied. (**d, g&j**). The number of the data points in the spectra is 288.

Finally, so far we focused on structures with link lengths satisfying $\beta L \cong m\pi$. However, the non-Hermitian skin effect still persists in asymmetrically coupled lasing arrays even when $\beta L$ significantly deviates from this condition (Fig. 4a). Figures 4b, c, f, and g display the intensity profiles and spectra of a two-level system under such phase locking conditions (see Supplementary Section 7 and 8 for detailed analysis). Even though in these cases the emission spectra involve two lasing modes, the intensity profile across the array nevertheless follows the same trend observed in their single mode counterparts—a result of incoherent superposition of the fields (intensity superposition) associated with various modes in each site. Similar behavior is also observed in a 5-element lattice (Figs. 4d, e, h, and i) (see Supplementary Section 9 for detailed analysis). In these cases, the eigen-frequencies are in general complex due to the coupling phase, despite the fact that the array does not feature periodic boundary conditions.

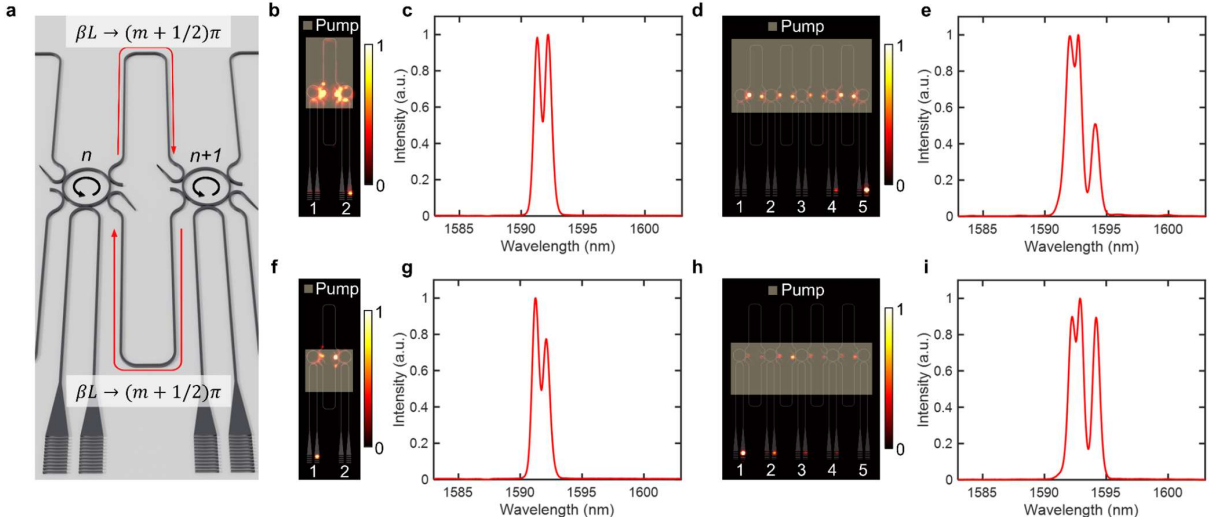

**Fig. 4 | Non-Hermitian skin effect in multi-mode laser arrays. a,** Schematic of the microring laser array where the length of the links is set to be central wavelength $\beta L \to (m + 1/2)\pi$ to observe multi-mode behavior. **b,c,f,g,** Multi-mode lasing operation in 2-element Hatano-Nelson system with the aforementioned link phase condition. The output intensity (incoherent superposition of two modes) is tilted towards right end (**b&c**) and left end (**f&g**), respectively. **d,e,h,i,** Multi-mode lasing behaviors in 5-element systems. The output intensity localizes at the right end (**d&e**) when the upper links are pumped, and left side (**h&i**) when the lower ones are illuminated. These results confirm that the non-Hermitian skin effect universally applies to all modes and their incoherent (intensity) superposition. The number of the data points in the spectra is 288.

## Discussion

In this work we focus on the simplest case of 1D Hatano-Nelson model. More complicated arrangements, for examples the ones that expand the demonstrated array to 2D and utilize high order hopping such as next nearest neighbor couplings, can pave the path towards large-scale phase-locked laser arrays with controllable intensity profiles and phase profiles. Theoretical works that involve similar concepts have proposed pulse-shortening in synthetic dimensions[38].

In conclusion, we have reported the first realization of Hatano-Nelson lattices in a laser setting by judiciously introducing a non-symmetric coupling between active resonators. The skin effect observed in such lattices may provide a new approach for near-field beam steering where the energy distribution throughout the array is globally controlled by modulating the gain/loss levels in the link areas. We also explored the effect of non-Hermitian delayed mutual coupling on laser phase locking. Our work may provide new avenues for near-field beam steering of phase locked laser arrays while shedding light on the intriguing physics of non-symmetrically coupled systems.

## Materials and Methods

The details of sample fabrication and experimental setup are provided in Supplementary Information section 2 and 4, respectively.

## Acknowledgements

We gratefully acknowledge the financial support from Air Force Office of Scientific Research (AFOSR) Multidisciplinary University Research Initiative (MURI) Award on Novel light-matter interactions in topologically non-trivial Weyl semimetal structures and systems (Award No. FA9550-20-1-0322), AFOSR MURI Award on Programmable systems with non-Hermitian quantum dynamics (Award No. FA9550-21-1-0202), Office of Naval Research (ONR) (N00014-19-1-2052, N00014-20-1-2522), ONR MURI award on the classical entanglement of light (Award No. N00014-20-1-2789), DARPA (D18AP00058), National Science Foundation (CBET 1805200), W. M. Keck Foundation, US–Israel Binational Science Foundation (BSF: 2016381), MPS Simons collaboration (Simons grant 733682), US Air Force Research Laboratory (FA86511820019) and the Qatar National Research Fund (grant NPRP13S0121-200126). G. G. Pyrialakos would like to acknowledge the support from Bodossaki Foundation. The authors also acknowledge the technical support from Dr. Andrew Wilkey in preparing the manuscript.

## Author contributions

M.K. and D.N.C. conceived the idea. Y.G.N.L and M.K. designed the structures and experiments. Y.G.N.L and O.H. fabricated and characterized the samples. Y.G.N.L, Y.W and O.H. performed the simulations. Y.G.N.L, Y.W, G.G.P, P.S.J, D.N.C and M.K. developed the theoretical models and analyzed the data. All authors contributed in preparing the manuscript.

## Competing interests

The authors declare no competing interests.

## Data availability

All data needed to evaluate the conclusions in the paper are present in the main text and methods. The datasets generated during and analyzed during this study are available from the corresponding author on reasonable request.

## Supplementary Information

Supplementary information accompanies the manuscript on the Light: Science & Applications website (http://www.nature.com/lsa)

**References:**

1. Hatano, N. & Nelson, D. R. Localization transitions in non-Hermitian quantum mechanics. *Physical Review Letters* **77**, 570-573 (1996).

2. Hatano, N. & Nelson, D. R. Vortex pinning and non-Hermitian quantum mechanics. *Physical Review B* **56**, 8651-8673 (1997).

3. Hatano, N. & Nelson, D. R. Non-Hermitian delocalization and eigenfunctions. *Physical Review B* **58**, 8384-8390 (1998).

4. Gong, Z. P. *et al*. Topological phases of non-Hermitian systems. *Physical Review X* **8**, 031079 (2018).

5. Yao, S. Y. & Wang, Z. Edge states and topological invariants of non-Hermitian systems. *Physical Review Letters* **121**, 086803 (2018).

6. Okuma, N. *et al*. Topological origin of non-Hermitian skin effects. *Physical Review Letters* **124**, 086801 (2020).

7. Kawabata, K., Sato, M. & Shiozaki, K. Higher-order non-Hermitian skin effect. *Physical Review B* **102**, 205118 (2020).

8. Xiao, L. *et al*. Non-Hermitian bulk–boundary correspondence in quantum dynamics. *Nature Physics* **16**, 761-766 (2020).

9. Kunst, F. K. *et al*. Biorthogonal bulk-boundary correspondence in non-Hermitian systems. *Physical Review Letters* **121**, 026808 (2018).

10. Zhang, L. *et al*. Acoustic non-Hermitian skin effect from twisted winding topology. *Nature Communications* **12**, 6297 (2021).

11. Yokomizo, K. & Murakami, S. Non-bloch band theory of non-Hermitian systems. *Physical Review Letters* **123**, 066404 (2019).

12. Weidemann, S. *et al*. Topological funneling of light. *Science* **368**, 311-314 (2020).

13. Wang, K. *et al*. Generating arbitrary topological windings of a non-Hermitian band. *Science* **371**, 1240-1245 (2021).

14. Wang, K. *et al*. Topological complex-energy braiding of non-Hermitian bands. *Nature* **598**, 59-64 (2021).

15. El-Ganainy, R. *et al*. Non-Hermitian physics and PT symmetry. *Nature Physics* **14**, 11-19 (2018).

16. Ren, J. H. *et al*. Unidirectional light emission in PT-symmetric microring lasers. *Optics Express* **26**, 27153-27160 (2018).

17. Zhong, Q. *et al*. Sensing with exceptional surfaces in order to combine sensitivity with robustness. *Physical Review Letters* **122**, 153902 (2019).

18. Liu, Y. G. N. *et al*. Engineering interaction dynamics in active resonant photonic structures. *APL Photonics* **6**, 050804 (2021).

19. Song, F., Yao, S. Y. & Wang, Z. Non-Hermitian skin effect and chiral damping in open quantum systems. *Physical Review Letters* **123**, 170401 (2019).

20. Li, L. H., Lee, C. H. & Gong, J. B. Topological switch for non-Hermitian skin effect in cold-atom systems with loss. *Physical Review Letters* **124**, 250402 (2020).

21. Lee, C. H., Li, L. H. & Gong, J. B. Hybrid higher-order skin-topological modes in nonreciprocal systems. *Physical Review Letters* **123**, 016805 (2019).

22. Longhi, S. Selective and tunable excitation of topological non-Hermitian quasi-edge modes. *Proceedings of the Royal Society A*: *Mathematical*, *Physical and Engineering Sciences* **478**, 20210927 (2022).

23. Longhi, S., Gatti, D. & Valle, G. D. Robust light transport in non-Hermitian photonic lattices. *Scientific Reports* **5**, 13376 (2015).

24. Nasari, H. *et al*. Observation of chiral state transfer without encircling an exceptional point. *Nature* **605**, 256–261 (2022).

25. Schumer, A. *et al.* Topological modes in a laser cavity through exceptional state transfer. *Science* **375**, 884–888 (2022).

26. Nelson, D. R. & Shnerb, N. M. Non-Hermitian localization and population biology. *Physical Review E* **58**, 1383-1403 (1998).

27. Amir, A., Hatano, N. & Nelson, D. R. Non-Hermitian localization in biological networks. *Physical Review E* **93**, 042310 (2016).

28. Kawabata, K. *et al*. Symmetry and topology in non-Hermitian physics. *Physical Review X* **9**, 041015 (2019).

29. Brandenbourger, M. *et al*. Non-reciprocal robotic metamaterials. *Nature Communications* **10**, 4608 (2019).

30. Ghatak, A. *et al*. Observation of non-Hermitian topology and its bulk–edge correspondence in an active mechanical metamaterial. *Proceedings of the National Academy of Sciences of the United States of America* **117**, 29561-29568 (2020).

31. Helbig, T. *et al*. Generalized bulk–boundary correspondence in non-Hermitian topolectrical circuits. *Nature Physics* **16**, 747-750 (2020).

32. Liu, S. *et al*. Non-hermitian skin effect in a non-Hermitian electrical circuit. *Research* **2021**, 5608038 (2021).

33. Scheibner, C., Irvine, W. T. M. & Vitelli, V. Non-Hermitian band topology and skin modes in active elastic media. *Physical Review Letters* **125**, 118001 (2020).

34. Guddala, S. *et al*. All-optical nonreciprocity due to valley polarization pumping in transition metal dichalcogenides. *Nature Communications* **12**, 3746 (2021).

35. Hu, X. X. *et al*. Noiseless photonic non-reciprocity via optically-induced magnetization. *Nature Communications* **12**, 2389 (2021).

36. Kim, S. *et al*. On-chip optical non-reciprocity through a synthetic Hall effect for photons. *APL Photonics* **6**, 011301 (2021).

37. Liu, Y. G. N. *et al*. Gain-induced topological response via tailored long-range interactions. *Nature Physics* **17**, 704-709 (2021).

38. Yuan, L. Q. *et al*. Pulse shortening in an actively mode-locked laser with parity-time symmetry. *APL Photonics* **3**, 086103 (2018).